\documentclass[lettersize,journal]{IEEEtran}
\usepackage{amsmath,amsfonts}
\usepackage{algorithmic}
\usepackage{algorithm}
\usepackage{array}
\usepackage[caption=false,font=normalsize,labelfont=sf,textfont=sf]{subfig}
\usepackage{textcomp}
\usepackage{stfloats}
\usepackage{url}
\usepackage{verbatim}
\usepackage{graphicx}
\usepackage{cite}
\usepackage{lineno,hyperref}
\usepackage{hyperref}
\usepackage{amsmath}
\usepackage{graphicx}
\usepackage{color}
\usepackage{amsthm,amsmath,amssymb}
\usepackage{bm}
\usepackage{mathrsfs}
\usepackage{booktabs} 
\usepackage{lineno}
\usepackage{multicol}
\usepackage{multirow}
\usepackage{lipsum}  
\usepackage{lineno}
\begin{document}
\title{Vaccination dynamics of age-structured populations in higher-order social networks}

\author{Yanyi Nie, Tao Lin, Yanbing Liu, Wei Wang	
\thanks{This work was supported in part by the Social Science Foundation of Chongqing (No. 2021PY53), the Natural Science Foundation of Chongqing (No. cstc2021jcyj-msxmX0132), the Natural Science Foundation of Yuzhong District, Chongqing (No. 20210117), the Science and Technology Research Program of Chongqing Municipal Education Commission (No. KJQN202200429), Program for Youth Innovation in Future Medicine, Chongqing Medical University (No. W0150), and National Natural Science Foundation of China under Grants (No. 62272074).\\(\emph{Corresponding author: Tao Lin and Wei Wang.})}
\thanks{Yanyi Nie and Tao Lin are with the College of Computer Science, Sichuan University, Chengdu 610065, China (email: nienienie@stu.scu.edu.cn; lintao@scu.edu.cn).}
\thanks{Yanbing Liu is with the Chongqing Medical University, Chongqing, 400016, China (email: liuyb@cqupt.edu.cn).}
\thanks{Wei Wang is with the School of Public Health, Chongqing Medical University, Chongqing, 400016, China (email: wwzqbx@hotmail.com and wwzqbc@cqmu.edu.cn).}}



\maketitle

\begin{abstract}
Voluntary vaccination is essential to protect oneself from infection and suppress the spread of infectious diseases. Voluntary vaccination behavior is influenced by factors such as age and interaction patterns. Differences in health consciousness and risk perception based on age result in heterogeneity in vaccination behavior among different age groups. Higher-order interactions among individuals of various ages facilitate the dissemination of vaccine-related information, further influencing vaccination intentions. To investigate the impact of individual age and interaction patterns on vaccination behavior, we propose an epidemic-game coevolution model in which age structure and higher-order interactions are considered. Combining the theoretical analysis of epidemic-game coevolution, this work calculates the evolutionarily stable strategies and dynamic equilibrium based on imitation dynamics in the well-mixed population. Extensive numerical experiments show that infants and the elderly exhibit conservative attitudes towards vaccination, and the vaccination levels of these two groups have no significant impact on the vaccination behavior of other age groups. The vaccination behavior of children is highly active, while the vaccination behavior of adults depends on the relative cost of vaccination. The increase in vaccination levels among children and adults leads to a decrease in vaccination levels in other groups. Furthermore, the infants exhibit the lowest level of vaccination, while the children have the highest vaccination rate. Higher-order interactions significantly enhance vaccination levels among children and adults.
\end{abstract}


\section{Introduction}
The global outbreak of COVID-19 has disrupted the order in all countries and communities, posing a significant threat to human life and property \cite{Brulhart2021Mental,Zhao2021Binary,Tayarani-Najaran2022Novel}. Scholars found that voluntary vaccination is the primary public health measure to intervene and control infectious diseases \cite{Chen2020Global,Dai2021Behavioural}. In particular, the emergence of variant coronaviruses (e.g., Omicron) due to pathogen mutations has driven the provision of booster doses for COVID-19 vaccines. Voluntary vaccination decisions are influenced by various factors, such as vaccine cost, peer decisions, personal experiences, and information dissemination from public health institutions \cite{Zuo2023Exploring,Huang2021Integrated}. Age, in particular, is an important factor affecting vaccination behavior \cite{Kovaiou2007Agerelated}. Differences in individual immune response, social interactions, infection risks, and vaccine efficacy among individuals of different ages contribute to the heterogeneity of vaccination behavior across age groups. The Ref.~\cite{Hethcote1999Simulations}, for example, indicated that increasing adult pertussis booster vaccinations every ten years helps reduce the incidence rate in adults, but only leads to a slight decrease in the rates for infants and toddlers. Additionally, in countries such as Austria, France, and Portugal, the interval for booster shots for individuals aged 65 and above is shortened, as antibody levels decline more rapidly with age \cite{Ciabattini2018Vaccination}.

The collective outcome of individual vaccination behavior determines the level of population immunity. Herd immunity is achieved when the vaccination coverage is sufficiently high, thereby eradicating the infectious diseases \cite{Bauch2004Vaccinationa,Khajanchi2017Role,Sarkar2023Spatiotemporala}. However, the high transmissibility of the pathogen poses challenges to achieving herd immunity through voluntary vaccination \cite{Galvani2007Longstandingc,Vardavas2007Cana}. Besides, economic costs, logistical constraints, religious beliefs, and other sociocultural factors often lead to difficulties in vaccination implementation \cite{Bauch2004Vaccinationa}. Developing personalized vaccination strategies based on different age groups, such as mass vaccination with transmission-blocking vaccines for adults \cite{Bubar2021Modelinformed} or prioritizing vaccine distribution to the elderly, can improve vaccine effectiveness and suppress infection to some extent \cite{Zhou2019Globala}. Specifically, in early 2021, the Singapore government implemented a vaccine rollout plan based on prioritization. Healthcare workers, frontline personnel, and other essential workers were identified as the priority groups, followed by the elderly aged 70 and above \cite{Zhang2022role}. This initiative significantly enhanced the effectiveness of vaccination and reduced mortality rates. Bubar et al. \cite{Bubar2021Modelinformed} found that prioritizing efficient transmission-blocking vaccination for adults aged 20-49 can maximally reduce cumulative incidence rates. In situations where vaccine resources are scarce, allocating vaccines to the population aged 20 to 59 is superior to the traditional ``seasonal" vaccination strategy targeting high-risk groups such as the elderly and children \cite{Chowell2009Adaptive}.

The dynamics of infectious diseases spread in a population fundamentally depend on the patterns of interaction\cite{Liu2021InfectionProbabilityDependent,Hong2022Personalized,Zhan2018Epidemic}, further influencing vaccination behavior \cite{Wang2016Identifying,Feehan2021Quantifying,Khajanchi2017Modeling}. For instance, the interaction patterns occurring in schools, conferences, and workplaces differ in frequency and distance, which yields different vaccination outcomes \cite{Wang2021Epidemica,Miyoshi2021Flexible}. The occurrence of contacts between three or more nodes is described as higher-order interactions \cite{Sun2022Diffusiona,Gao2017NetworkBaseda}, which can happen in various real-life scenarios. For example, collective discussions in conference rooms, dining together in restaurants, or group training on playgrounds all involve a group of people engaging in collective behavior within the same setting. Higher-order interactions facilitate the dissemination of information, such as vaccination information, which can have a significant impact on vaccination behavior. Furthermore, previous research has shown that the presence of higher-order interactions profoundly affects the dynamics of the infectious process and can give rise to phenomena such as bistability, hysteresis, and explosive transitions \cite{Battiston2021physicsj}.

Specifically, the household setting is an important scenario where higher-order interactions occur. In particular, the implementation of non-pharmaceutical interventions \cite{Chen2014Optimal,Hong2023Integrated}, such as travel restrictions and mandatory stay-at-home orders, has greatly facilitated higher-order interactions among family members, leading to a significant portion of infections occurring within households \cite{Pangallo2024unequal}.

In order to capture higher-order interactions, relying solely on simple graphs to characterize them can lead to drawbacks such as information loss \cite{Battiston2021physicsj}. Therefore, more advanced mathematical structures, such as hypergraphs, simplicial complexes and motifs, have been proposed to represent systems with higher-order interactions. A hypergraph is defined by a node set $\mathcal{V}$ and a set of hyperedges $\mathcal{H}$, where these hyperedges $\mathcal{H}$ specify which nodes participate and how in interactions \cite{Battiston2020Networksj,St-Onge2022Influential,Kim2024HigherOrder}. Network motifs are the most common higher-order structures, defined as specific patterns of edges between vertices that appear statistically significant in the network \cite{Battiston2020Networksj}. The higher-order organizational structures manifested in networks based on specific social, age, or occupational relationships can be effectively captured using the motifs \cite{Benson2016Higherorderd}. In particular,
simplicial complexes, as suitable tools capable of capturing both pairwise interactions and higher-order interactions simultaneously, satisfy the additional constraint that any subset of a simplex is also a simplex \cite{Iacopini2019Simplicialg,Zhang2023Impact}.
In recent years, simplicial complexes have been extensively studied and applied in various research fields, such as collaborative networks, semantic networks, cellular networks, and brain networks \cite{Sun2022Diffusiona,Sizemore2018Knowledge,Lee2012Persistent}. To characterize a complex contagion process on simplicial complexes, a simplicial contagion model was recently developed \cite{Iacopini2019Simplicialg}. 
Based on the susceptible-infected-susceptible (SIS) model \cite{Sanatkar2016Epidemic}, 
Li et al. \cite{Li2021Contagiona} captured coexistence of interactions with different orders and found that higher-order simplices drive the dynamics of infection. Focusing on the spreading dynamics on the simplicial susceptible-infected-recovered (s-SIR) model, Palafox-Castillo et al. \cite{Palafox-Castillo2022Stochastic} defined a stochastic model to study variations beyond contagion processes on simplicial networks. 

Based on the above analysis, as emphasized by Rosas et al. \cite{Rosas2022Disentangling}, the study of higher-order phenomena should focus on both the higher-order interdependencies inferred from observed data and the use of topological data analysis to examine the structure of the system. This approach allows for a more comprehensive understanding of the mechanisms and behaviors of higher-order systems. Mechanisms provide the ``underlying rules" of epidemic spread and offer the foundational parameters for mathematical modeling. Behaviors, on the other hand, represent the expression of these mechanisms at the population level, revealing the macro patterns and trends of disease transmission, which are essential for assessing the model's accuracy and real-world applicability. In our model, simplicial complexes determine the basic rules and structure of epidemic spread. Additionally, by simulating and analyzing behaviors, we can assess the development trends of the outbreak and validate the effectiveness of preventive measures.

Evolutionary game theory, as a mathematical analysis tool, is well suited to be combined with epidemiological models \cite{Gao2013Modeling,Gao2022Epidemic} to describe the vaccination dynamics. 
In recent years, human behavioral factors such as imitation and free-riding have been found to significantly influence the ultimate vaccination levels. Zuo et al. \cite{Zuo2023Exploring}, for example, established a cost-benefit analysis function that represents vaccine decision-making, and found that vaccination information provides a favorable reference in individuals' decision-making regarding vaccine uptake. Wang et al. \cite{Wang2020Vaccination} constructed a two-layer network regarding vaccine decision-making and imitation behavior and they found that imitation behavior inhibits the improvement of herd immunity. The ``role model" groups within peers also influence vaccination behavior. Bauch et al. \cite{Fu2011Imitationb} proposed an evolutionary game theory model to reveal how imitation by peers influences vaccination decisions. The experimental results demonstrate  that committed vaccine uptake by individuals can effectively stimulate vaccination behavior among peers \cite{Liu2012Impact}. 
Considering that awareness is an important factor influencing vaccination, Kabir et al. \cite{Kabir2019Effect} proposed a framework for vaccine uptake with the unaware-aware (UA) information propagation. They found that the spread of information can suppress infectious disease transmission.


The issue of the vaccination dilemma is a long-standing challenge. Existing research has extensively explored the impact of external factors such as information dissemination and peer influence on vaccination. However, due to variations in internal factors such as age and interaction patterns, notable differences exist in individual immunity, vaccine efficacy, and vaccine information acquisition \cite{Yamin2013Incentives}, leading to heterogeneity in vaccination. Neglecting the heterogeneity and devising the same vaccination mechanism for all individuals may make it difficult to capture the real dynamics of vaccination. Our work aims to characterize the vaccination characteristics of different age groups and explore the impact of age and interaction patterns on vaccination. This is intended to help build more accurate models and provide valuable information for the development of vaccine promotion strategies. Therefore, the work is motivated by the following three aspects:
(1) Population age is a crucial reference for formulating vaccination strategies. However, the differences in vaccination behavior resulting from population age are still unclear.
(2) Interaction patterns significantly influence individual vaccination behavior and the outbreak of infectious diseases. However, the impact of higher-order interactions, prevalent in social scenarios, on vaccination intentions remains to be investigated.
(3) There is still a lack of relevant theoretical foundation for exploring the influence of internal factors such as contact patterns and age on individual vaccination behavior.
%
%
%

The work constructs an epidemic-game coevolution model to characterize the vaccination characteristics of different age groups and explore the impact of age and interaction patterns on vaccination.
We calculate the evolutionarily stable strategies (ESSs) and dynamic equilibrium based on imitation dynamics in the well-mixed population.
Extensive experimental results indicate conservative vaccination characteristics in baby and elderly, highly active vaccination behavior in child, and vaccination behavior in adult depending on the relative vaccination cost. 
In addition, higher-order interactions led to a significant increase in child and adult vaccination levels. In general, this work focuses on individual heterogeneity among different age groups, revealing distinct characteristics in vaccine attitudes and behaviors among infants, children, adults and elderly. Our work also introduces the concept of higher-order interactions, reflecting the complexity of real-life interpersonal interactions. By revealing interaction patterns among individuals and understanding individual heterogeneity, the work provides insights into vaccination dynamics, which is of practical significance for devising targeted vaccine promotion strategies and optimizing infectious disease control measures.

In the following, we first describe the details of our model in Sec.~\ref{model}. Next, we develop an analytical framework for the epidemic-game coevolution model with the age structure by using a mean-field approach in Sec.~\ref{theoretical}. In Sec.\ref{numerical}, numerical experiments are implemented to capture the behavioral characteristics of vaccination in various age groups. Finally, we present some discussions and conclude the paper in Sec.~\ref{conclusion}.

\section{Model Structure}\label{model}
\begin{figure*}[h]
	\centering
	\includegraphics[scale=0.62]{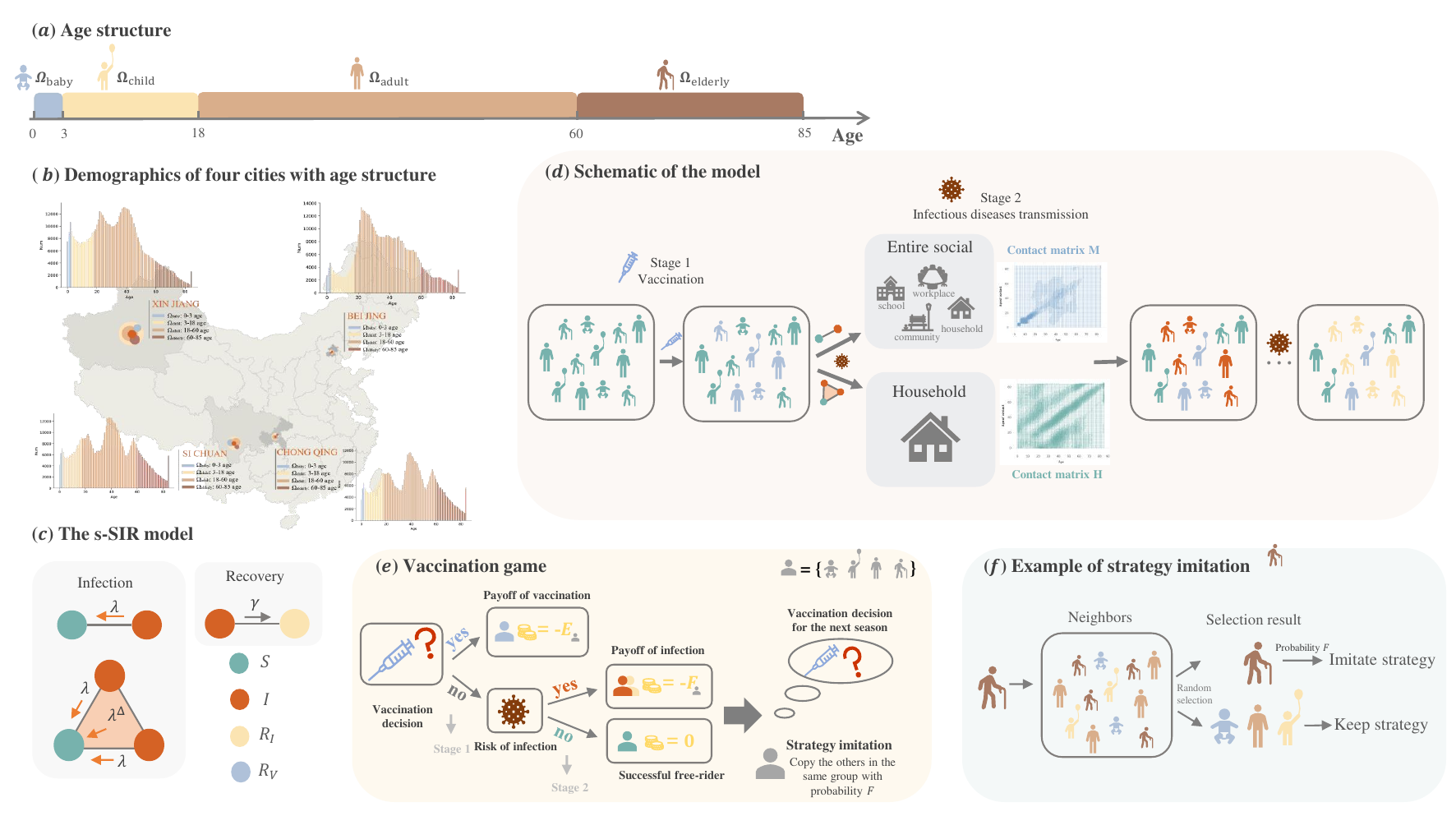}
	\caption{(Color online) Schematic illustration of epidemic-game coevolution model based on the age structure. (a) Details of the division of the age structure. The population in the system is divided into $\Omega_{\rm{baby}}$, $\Omega_{\rm{child}}$, $\Omega_{\rm{adult}}$, and $\Omega_{\rm{elderly}}$ four groups. The demographics of the four cities based on the age structure are shown in (b). (c) The s-SIR contagion model with order up to $D=2$. (d) The epidemic-game coevolution model with the age structure. Each flu season covers two steps: vaccination (stage 1) and dual-scenario infectious disease transmission (stage 2). A well-mixed population is considered. The color changes of the icons of different shapes represent the state transition of the individual. (e) The two-step vaccination game with the age structure.  (f) Example of strategy imitation for people in $\Omega_{\rm{elderly}}$.}
	\label{fig1}
\end{figure*}
In this section, we propose an epidemic-game coevolution model covering the age structure (Sec.~\ref{age}), dual-scenario infection (Sec.~\ref{infection}), and public vaccination (Sec. ~\ref{vaccination}). In general, the preemptive vaccination strategy is adopted. Therefore, as shown in Figs.~\ref{fig1}(d), we specify that individuals of each age group decide whether to be vaccinated before the start of the seasonal infectious diseases. The dual-scenario infection then determines whether each susceptible individual, i.e., unvaccinated individuals, is infected at a certain time step during the influenza season. Once the epidemic ends, based on a vaccination game with the age structure, individuals in each age group decide the vaccination strategy for the next season until the system reaches equilibrium.


\subsection{Age structure}\label{age}

In our model, the individuals in the system are divided into $N$ age based on the age of the population. Considering that the age is not necessarily an integer, an individual whose age is $i\in \mathbb{A}(0,N)$, where $\mathbb{A}(0,N)=\{age|0\leq age<N,age\in \rm{Z}\}$, meaning that his/her actual age is between $i$ to $i+1$. We define $\bm{u}_i$ to denote an individual with age $i\in\mathbb{A}(0,N)$.

In reality, vaccination strategies tend not to be specific to each age, but are differentiated according to, for example, the age groups of the population: children, adults, and the elderly \cite{Boldea2022Agespecific}. Inspired by this fact, as shown Figs.~\ref{fig1}(a), we further divide the population into the following four groups according to their age.

$\bullet$ $\Omega_{\rm{baby}}$: individuals with age $i\in\mathbb{A}(0,A)$. This group is used to model real-world preschool populations of infants and toddlers. 

$\bullet$ $\Omega_{\rm{child}}$: individuals with age $i\in\mathbb{A}(A,B)$. The pre-adult population involved in schooling is described by group $\Omega_{\rm{child}}$. 

$\bullet$ $\Omega_{\rm{adult}}$: individuals with age $i\in\mathbb{A}(B,C)$. The adult population is characterized by $\Omega_{\rm{adult}}$. 

$\bullet$ $\Omega_{\rm{elderly}}$: individuals with age $i\in\mathbb{A}(C,N)$. This group corresponds to the elderly population. 

If individual $\bm{u}_i \in \Omega_{g}$, $g \in\{\rm{baby}, child, adult, elderly\}$, it means that the age $i$ of $\bm{u}_i$ fits the age range of the group $\Omega_g$. Based on the age structure, the Figs.~\ref{fig1}(b) show the demographics of each age and the four groups in the four Chinese cities.


In this work, the incorporation of age structure plays a crucial role in enhancing the authenticity and completeness of our model. Firstly, considering age structure helps ensure that the model aligns more closely with real-world conditions. The immune system, physiological status, and social behaviors vary with age, and incorporating age structure into the model allows for a more accurate reflection of different groups' infection and immune responses to vaccines. Secondly, considering age structure allows the model to reflect the vaccination behavior of the entire population more comprehensively. Combining age structure enables the analysis of heterogeneous behaviors among individuals in various age groups rather than being confined to a single age group.

\subsection{Dual-scenario infection}\label{infection}
Influenza outbreaks are particularly associated with interactions that occur in a variety of society scenarios, such as households, schools, workplaces, and general communities \cite{Mistry2021Inferring}. 
The household, is typically the scenario with the most frequent collective interactions and highest infection rates across all ages \cite{Pangallo2024unequal}. Therefore, based on the age structure, we construct dual-scenario infection process, which covers the entire society and the household scenario.

Here, we consider a simplicial susceptible-infected-removal (s-SIR) model, which is similar to the s-SIS model proposed in Ref.~\cite{Iacopini2019Simplicialg}. The s-SIR model is typically employed to describe epidemics whose evolution is irreversible. The irreversible epidemics, such as chickenpox, measles, whooping cough and seasonal influenza, grant individuals lifelong immunity once cured. This work uses seasonal influenza as a specific example to assist the study, but our model can still be effectively adapted to describe other SIR epidemics.

First, based on the age structure, we define four states, $\mathcal{S}$, $\mathcal{I}$, $\mathcal{R_I}$, $\mathcal{R_V}$, for individuals with age $i\in\mathbb{A}(0,N)$ in the model. 

$\bullet$ Susceptible state: $\mathcal{S}$. Individuals in the state $\mathcal{S}$ are susceptible to the virus and may contract infectious diseases during contact with others.

$\bullet$ Infected state: $\mathcal{I}$. Individuals are infected with the virus and have the ability to infect other individuals.

$\bullet$ Removal state: $\mathcal{R}$. Individuals in state $\mathcal{R}$ are immune to the virus. Two population groups, $\mathcal{R_I}$ and $\mathcal{R_V}$, are included. 

$\bullet$ $\mathcal{R_I}:$ After being infected, individuals recover from the state $\mathcal{I}$ and gain immunity. Such individuals no longer participate in the infection process.

$\bullet$ $\mathcal{R_V}:$ Individuals gain immunity through vaccination when they are healthy and are not involved in the infection process. 

In particular, at each time step, an individual $\bm{u}_i$ belongs to only one state. 

Besides, since vaccinated individuals are fully immune to the virus throughout the seasonal influenza, the infection process involves only three states, $\mathcal{S}$, $\mathcal{I}$, $\mathcal{R_I}$, of the individual. 

Next, we describe the infection and recovery process in dual-scenario, i.e., the state transition of individuals of different ages as shown in Figs.~\ref{fig1}(c). The contact matrices $\rm{\textbf{M}}$ and $\rm{\textbf{H}}$ are introduced to characterize interaction patterns in different scenarios (see the electronic supplementary materials).

$\bullet$ Pairwise contagion. In the entire society, an $\mathcal{S}$-state individual $\bm{u}_i$ and an $\mathcal{I}$-state individual $\bm{v}_j$ come into contact through pairwise interaction, the individual $\bm{u}_i$ transforms to the $\mathcal{I}$-state with probability $\beta$. Mathematically,
there is $$\mathcal{S}_i+\mathcal{I}_j\xrightarrow[infection]{M_{ij}\beta}\mathcal{I}_i+\mathcal{I}_j,$$ 
where the element $M_{ij}$ of matrix $\rm{\textbf{M}}$ describes pairwise interaction as described in the supplementary materials-A.

$\bullet$ Higher-order contagion. In the household scenario, when the $\mathcal{S}$-state individual $\bm{u}_i$ is in contact with both the $\mathcal{I}$-state individual $\bm{v}_j$ and $\mathcal{I}$-state individual $\bm{x}_l$, through higher-order interactions, the individual $\bm{u}_i$ is subjected to reinforcement infection effect from the household scenario and shifts to the $\mathcal{I}$-state with probability $\beta^{\Delta}$. In particular, the ages $i$, $j$, and $l$ of individuals can be equal. The process is written as $$\mathcal{S}_i+\mathcal{I}_j+\mathcal{I}_l\xrightarrow[infection]{H_{ijl}\beta^{\Delta}}\mathcal{I}_i+\mathcal{I}_j+\mathcal{I}_l,$$
where $H_{ijl}$ is calculated from the joint distribution probabilities of the pairwise interaction, describing the higher-order interactions, as detailed in Supplementary Material-A. $H_{ijl} \beta^{\Delta}$ represents the probability of infection within the household scenario under the quenched mean-field assumption \cite{Ghosh2023Dimensionb,Li2024Social,Iacopini2019Simplicialg}. In our model, the infection is restricted to three individuals, i.e., the order of the s-SIR contagion model is up to $D=2$, as detailed in the supplementary materials-A \cite{Iacopini2019Simplicialg}.

$\bullet$ Recovery. When a individual is in $\mathcal{I}_i$-state, it spontaneously transitions to the $\mathcal{R_I}_i$-state with probability $\gamma$. There is
$$\mathcal{I}_i\xrightarrow[recovery]{\gamma}\mathcal{R_I}_i.$$

\subsection{Vaccination game with the age structure}\label{vaccination}
Vaccination dynamics are modeled as a two-stage game, as shown in Figs.~\ref{fig1}(e), i.e., public vaccination (referred to as stage 1) and infectious disease transmission (referred to as stage 2). These two stages are described in detail as follows.

$\bullet$ \textbf{stage 1}: Stage 1 is public vaccination, which occurs before any infection. At this stage, each $\mathcal{S}$-state individual $\bm{u}_i$ decides whether or not to be vaccinated. If individual an $\bm{u}_i$ is vaccinated, $\bm{u}_i$ transforms from the $\mathcal{S}$-state to the $\mathcal{R_V}$-state, which is described as
$$\mathcal{S}_i\xrightarrow[vaccination]{1}\mathcal{R_V}_i.$$ 
In addition, vaccinations incur vaccination costs. Specifically, an individual $\bm{u}_i \in \Omega_g$, $g \in\{\rm{baby}, child, adult, elderly\}$ who is vaccinated will incur vaccination cost $E_g$. The cost of vaccination $E_g$ includes direct monetary costs, adverse health effects from vaccines, etc. For simplicity, we assume that once individuals are vaccinated, they are completely immune to the infectious virus for that season. The vaccine is only effective for one season due to pathogen mutations and decreased immunity. The initial state consists of an equal proportion of vaccinated and non-vaccinated individuals. Vaccinated individuals are randomly distributed throughout the population. 

$\bullet$ \textbf{Stage 2}: Stage 2 is the infectious disease transmission. Assuming that the infection begins in a population of any age $i\in\mathbb{A}(0,N)$, i.e., the initial number of infections is $I_i(0)$. During the spread of seasonal influenza, according to the simplicial contagion model described in Sec.~\ref{infection}, a non-vaccinated one may be infected or recover. The infectious diseases spread until there are no new infections. Depending on the infection results in stage 2, the individual becomes a free-rider, or a new cost is incurred. Specifically, 

1) Non-vaccinated individual $\bm{u}_i$ is infected in the stage 2. The $\bm{u}_i \in \Omega_{g}$, $g \in\{\rm{baby}, child, adult, elderly\}$ needs to pay for the cost $F_g$ of infection.

2) Non-vaccinated individual $\bm{u}_i$ is still healthy in the stage 2. Individual $\bm{u}_i$ declined the vaccination and successfully avoided infection at no cost to him/her. Benefit from the vaccination efforts of others and become a lucky free-rider.

In addition, to scale the cost of vaccination and the cost of infection, define $C_g =\frac{E_g}{F_g}$ ($0<C_g<1$), $g \in\{\rm{baby}, child, adult, elderly\}$ as the relative vaccination cost. The $C_g$ value for modeling specific diseases can be estimated through surveys on health attitudes, behaviors, and outcomes, as indicated in Ref.~\cite{Galvani2007Longstandingc}. However, in general, the cost of vaccination is expected to be relatively low compared to the cost of infection \cite{Fu2011Imitationb}. Based on the principle that higher costs lead to lower benefits, the payoff $P_g^{\mathcal{R_V}}$ of vaccinated ($\mathcal{R_V}$-state) individuals in age group $\Omega_g$ can be calculated as $P_g^{\mathcal{R_V}}=-\frac{E_g}{F_g}=-C_g$. Similarly, $P_g^{\mathcal{S}}=0$ is the payoff of unvaccinated and healthy ($\mathcal{S}$-state) individuals in group $\Omega_g$, and $P_g^{\mathcal{I}}=-\frac{F_g}{F_g}=-1$ is the payoff of unvaccinated and infected ($\mathcal{R_I}$-state) individuals in group $\Omega_g$. Table~\ref{table1} shows payoffs $P_g^{\mathcal{A}}$ for individuals of age group $\Omega_g$ in state $\mathcal{A}$, with $\mathcal{A} \in \{\mathcal{S},\mathcal{R_I},\mathcal{R_V}\}$, where N/A indicates no infected individuals whose strategy is vaccination.

\begin{table}
	\centering
	\caption{The payoff for individual}
	\label{table1}
	\begin{tabular}{ccc}
		\toprule
		Strategy/state & Healthy & Infected  \\
		\midrule
		Vaccination &  $P_g^{\mathcal{R_V}}=-C_g$ & N/A  \\
		Non-vaccination & $P_g^{\mathcal{S}}=0$ & $P_g^{\mathcal{R_I}}=-1$ \\
		\bottomrule
	\end{tabular}
\end{table}

Once the epidemic ends, each individual decides on a vaccination strategy for the next season based on current gains. In our work, imitation dynamics \cite{Hofbauer1998Evolutionary} is utilized to study the evolution of individual strategies. Specifically, individual $\bm{u}_i$ randomly selects a neighbor node $\bm{v}_j$, i.e., any node connected to node $\bm{u}_i$. If individuals $\bm{u}_i$ and $\bm{v}_j$ belong to the same age group $\Omega_{g}$, $g \in\{\rm{baby}, child, adult, elderly\}$, $\bm{u}_i$ compares her own payoff to that of the $\bm{v}_j$. This assumption is based on the fact that, under the effect of homophily \cite{Hiraoka2022Herda}, individuals tend to imitate the behavior of their peers. The probability of individual $\bm{u}_i$ adopting the strategy of node $\bm{v}_j$ is given by the Fermi function 
\begin{equation}
	\label{Femi}
	F(P_{\bm{u}_i}^\mathcal{A},P_{\bm{v}_j}^\mathcal{A})=\frac{1}{1+e^{-\delta(P_{\bm{}_j}^\mathcal{A}-P_{\bm{u}_i}^\mathcal{A})}},
\end{equation}
where $\mathcal{A} \in\{\mathcal{S},\mathcal{R_I},\mathcal{R_V}\}$ and $\delta$ denotes the intensity of selection ( $0<\delta <\infty$). The Figs.~\ref{fig1}(f) illustrates the strategy imitation process using the people in group $\Omega_{\rm{elderly}}$ as an example.
\section{Mean-field equations}\label{theoretical}
In this section, we aim to mathematically model the dynamics of seasonal infectious disease transmission and the vaccination game based on the age structure. Under the well-mixed assumption, we introduce a mean-field (MF) method [57], which assumes that individuals of the same age are equivalent. In addition, the contact matrices $\rm{\textbf{\textsc{M}}}$ and $\rm{\textbf{H}}$ are used to describe pairwise and higher-order contact patterns, respectively. 

\subsection{Aged-structured SIR model}
Define the fraction of $\mathcal{S}$-state, $\mathcal{I}$-state, $\mathcal{R_I}$-state and $\mathcal{R_V}$-state individuals of age $i$ at time $t$ as $S_i(t)$, $I_i(t)$, $R_{I_i}(t)$ and $R_{V_i}(t)$, respectively. According to the s-SIR model in dual-scenario introduced in Sec.~\ref{infection}, the evolution equations for the population state are
\begin{equation}
	\label{eq_2}
	\begin{aligned}
		\frac{dS_i(t)}{dt}=&-S_i(t)\beta \sum_j M_{ij} I_j(t)\\&-S_i(t)\frac{\beta^{\Delta}}{2}\sum_j \sum_l H_{ijl}I_j(t)I_l(t),
	\end{aligned}	
\end{equation}
\begin{equation}
	\label{eq_3}
	\begin{aligned}
		\frac{dI_i(t)}{dt}=&-\gamma I_i(t)+S_i(t)\beta \sum_j M_{ij} I_j(t)\\&+S_i(t)\frac{\beta^{\Delta}}{2}\sum_j \sum_l H_{ijl}I_j(t)I_l(t),
	\end{aligned}	
\end{equation}
\begin{equation}
	\label{eq_4}
	\begin{aligned}
		\frac{dR_{I_i}(t)}{dt}=&\gamma I_i(t),
	\end{aligned}	
\end{equation}
where $\beta$, $\beta^{\Delta}$ and $\gamma$ represent the infection rate of pairwise interaction in the entire social scenario, reinforcement infection effect from the household scenario, and recovery rate, respectively. The $M_{ij}$ represents the average number of contacts with individuals of age $j$ for an individual of age $i$ in the entire social scenario. Similarly, $H_{ijl}$ is the average number of times an individuals aged $i$ who are exposed to both individuals of age $j$ and $l$ in the household scenario.

Here, consider that pairwise contagions constitute the predominant pathway leading to the final epidemic size. This is because the number of interactions that occur between any two individuals in a social scenario is often greater than the number of interactions they have in the household scenario \cite{Pangallo2024unequal}. Therefore, in order to analytically calculate the final epidemic size and for simplicity, we can ignore the higher-order terms in Eqs.~(\ref{eq_2})-(\ref{eq_4}) and consider only the special form with pairwise contagions as
\begin{equation}
	\label{eq_5}
	\begin{aligned}
		\frac{dS_i(t)}{dt}=&-S_i(t)\beta \sum_j M_{ij} I_j(t),
	\end{aligned}	
\end{equation}
\begin{equation}
	\label{eq_6}
	\begin{aligned}
		\frac{dI_i(t)}{dt}=&-\gamma I_i(t)+S_i(t)\beta \sum_j M_{ij} I_j(t),
	\end{aligned}	
\end{equation}
\begin{equation}
	\label{eq_7}
	\begin{aligned}
		\frac{dR_{I_i}(t)}{dt}=&\gamma I_i(t).
	\end{aligned}	
\end{equation}

Denote $\mathscr{R}_{ij}=\frac{\beta}{\gamma}M_{ij}$ as the number of secondary infections in individuals of age $i$ caused by an infectious individual of age $j$ \cite{Pangallo2024unequal}. Then the Eq.~(\ref{eq_7}) can be rewritten as 
\begin{equation}
	\label{eq_8}
	\begin{aligned}
		\frac{d\sum_j \mathscr{R}_{ij} R_{I_j}(t)}{dt}=\sum_j  \beta M_{ij} I_j(t).
	\end{aligned}	
\end{equation}
Combining Eq.~(\ref{eq_5}) and Eq.~(\ref{eq_8}), we obtain
\begin{equation}
	\label{eq_9}
	\begin{aligned}
		\frac{dS_i(t)}{dt}=&-S_i(t) \sum_j \mathscr{R}_{ij} \frac{dR_{I_j}(t)}{dt}.
	\end{aligned}	
\end{equation}
Simplifying Eq.~(\ref{eq_9}), we have
\begin{equation}
	\label{eq_10}
	\begin{aligned}
		\frac{1}{S_i(t)}dS_i(t)=-\sum_j \mathscr{R}_{ij} dR_{I_j}(t).
	\end{aligned}	
\end{equation}
Integrating above equation from time $0$ to $\infty$, there are
\begin{equation}
	\label{eq.11}
	\int_{0}^{\infty}\frac{1}{S_i}dS=-\int_{0}^{\infty}\sum_j \mathscr{R}_{ij} dR_{I_j}(t).
\end{equation}
Then the transcendental equation for the final epidemic size of individuals with age $i$ can be obtained and written as 
\begin{equation}
	\label{eq.12}
	S_i(\infty)=S_i(0)e^{-\sum_j\mathscr{R}_{ij}[R_{I_j}(\infty)-R_{I_j}(0)]}.
\end{equation}
Consider the initial conditions of infectious disease caused by an infected person of age $i$, there are $S_i(0)\approx 1$ and $R_{I_i}(0)=0$. Using the initial condition, and the final state $S_i(\infty)=1-I_i(\infty)-R_{I_i}(\infty)=1-R_{I_i}(\infty)$, with $I_i(\infty)=0$, we obtain
\begin{equation}
	\label{eq.13}
	R_{I_i}(\infty)=1-e^{-\sum_j\mathscr{R}_{ij}R_{I_j}(\infty)},
\end{equation}
where $R_{I_i}(\infty)$ is the final fraction of individuals of age $i$ who had been infected during the infectious disease outbreak. In addition, the percentage of the infected population in different age structures $\Omega_g$, $g \in\{\rm{baby}, child, adult, elderly\}$ can be calculated as $R_{I_g}(\infty)=\frac{\sum_i R_{I_i}(\infty)}{i_{\rm{max}}-i_{\rm{min}}}$, where $i$ is the age for individuals in $\Omega_g$.

Consider preemptive vaccination with the age structure, i.e., assume that a proportion $x_g$ of individuals in $\Omega_g$, $g \in\{\rm{baby}, child, adult, elderly\}$ are vaccinated prior to an influenza outbreak. The final epidemic size can be rewritten according to the age structure as
\begin{equation}
	\label{eq.14}
	R_{I_g}(\infty)=(1-x_g)\frac{\sum_i R_{I_i}(\infty)}{i_{\rm{max}}-i_{\rm{min}}},
\end{equation}
where $i$ is the age of individuals in $\Omega_g$. Define $w(x_g)$ as the risk of infection for unvaccinated individuals in $\Omega_g$ when the proportion of vaccinated individuals in category $\Omega_g$ is $x_g$. So, the ratio of final epidemic size $R_{I_g}(\infty)$ with vaccination rate $x_g$  and the fraction of unvaccinated susceptible individuals $S_g(0)=1-x_g$ can define the risk of infection $w(x_g)$, written as
\begin{equation}
	\label{eq.15}
	w(x_g)=\frac{R_{I_g}(\infty)}{1-x_g}=\frac{\sum_i R_{I_i}(\infty)}{i_{\rm{max}}-i_{\rm{min}}},
\end{equation}
where

In addition, considering the vaccination game introduced in Sec.~\ref{vaccination} and infection risk $w(x_g)$, the Nash equilibrium in this game can be computed as 
	\begin{equation}
		\label{Nash}
		x_g^*=1+\frac{\mathrm{ln}(1-C_g)}{C_g\mathscr{R}_0},
	\end{equation}
which is a mixed strategy. The $\mathscr{R}_0$ is basic reproduction ratio for the entire population. Also, the Nash equilibrium is proven to be the evolutionarily stable strategy (ESS) (see the electronic supplementary materials-B).

\subsection{Evolution of vaccination behavior with the age structure} \label{evolution behavior}
Consider the well-mixed population $\Omega_g$ of size $N_g$, $g \in\{\rm{baby}, child, adult, elderly\}$, assume that the percentage of vaccinations is $x_g =\frac{V_g}{N_g}$. Recall the ``pairwise comparison rule" introduced in Sec.~\ref{vaccination}, whereby individuals prefer to imitate neighbors in the same age group with higher payoffs. Hence, the probability that the number of vaccinated individuals in group $\Omega_g$ increases from $V_g$ to $V_g+1$ is
\begin{equation}
	\label{eq.16}
	\begin{aligned}
		x_g^+&=x_g f_g (1-x_g)[1-w(x_g)]F(P_g^{S}, P_g^{R_V})\\&+x_g f_g (1-x_g)w(x_g)F(P_g^{I}, P_g^{R_V})\textcolor{red}{,}
	\end{aligned}
\end{equation}
where $f_g =\frac{N_g}{N}$ denotes the number of individuals in group $\Omega_g$ as a proportion of the total population. According to Eq.~(\ref{Femi}), the Fermi function $F(P_g^{S}, P_g^{R_V})$ describes the probability that an $\mathcal{S}$-state individual in group $\Omega_g$ imitates the strategy of an $\mathcal{R_V}$-state individual. Similarly, the probability that the number of vaccinated people decreases from $V_g$ to $V_g-1$ is
\begin{equation}
	\label{eq.17}
	\begin{aligned}
		x_g^-&=x_g f_g (1-x_g)[1-w(x_g)]F(P_g^{R_V}, P_g^{S})\\&+x_g f_g (1-x_g)w(x_g)F(P_g^{R_V}, P_g^{I}).
	\end{aligned}
\end{equation}

Taking into account Eqs.~\ref{eq.16} and \ref{eq.17}, the time evolution equation for $x_g$ can be calculated as 
\begin{equation}
	\label{eq.18}
	\begin{aligned}
		\frac{dx_g}{dt}&=x^+-x^-\\&
		=x_gf_g(1-x_g)\bigg([1-w(x_g)][F(P_g^{S},P_g^{R_V})-F(P_g^{R_V}, P_g^{S})]\\&
		+w(x)[F(P_g^{I},P_g^{R_V})-F(P_g^{R_V},P_g^{I})]\bigg)\textcolor{red}{.}
	\end{aligned}
\end{equation}
Substituting the Fermi function from Eq.~(\ref{Femi}) into Eq.~(\ref{eq.18}), we obtain
\begin{equation}
	\label{eq.19}
	\begin{aligned}
		\frac{dx_g}{dt}&=x_gu(1-x_g)\big([1-w(x_g)]\mathrm{tanh}[\frac{\delta}{2}(P_g^{\mathcal{R_V}}-P^{\mathcal{S}}_{g})]\\&+w(x_g) \mathrm{tanh} [\frac{\delta}{2}(P_g^{\mathcal{R_V}}-P_g^{\mathcal{R_I}})]\big),
	\end{aligned}
\end{equation}
which is the imitation dynamic equation. 
As shown in Table~\ref{table1}, the payoff of $\mathcal{A}$-state individuals in $\Omega_A$ has been defined, where $\mathcal{A} \in\{\mathcal{S},\mathcal{R_I},\mathcal{R_V}\}$. Taken together, the Eq.~(\ref{eq.19}) can be rewritten as
\begin{equation}
	\label{eq.20}
	\begin{aligned}
		\frac{dx_g}{dt}&=x_gf_g(1-x_g)\bigg([1-w(x_g)]\mathrm{tanh} (-\frac{\delta}{2}C_g)\\&+w(x_g)\mathrm{tanh}[\frac{\delta}{2}(1-C_g)]\bigg).
	\end{aligned}
\end{equation}
Utilizing the aforementioned system of equations, the final infection size and the vaccination level at equilibrium can be iteratively computed.
\subsection{Dynamic equilibrium analysis}

In this section, by considering different selection intensities, we analyze the dynamic equilibrium  \cite{Hofbauer1998Evolutionary,Khajanchi2017Uniform} based on imitation dynamics. The equilibria of Eq.~(\ref{eq.20}) can be categorized as follows.


	$\bullet$ Case 1: The intensity of selection $\delta \rightarrow 0$. So we have $\mathrm{tanh}(\delta x) \sim \delta x$. The Eq.~(\ref{eq.20}) simplifies to
	\begin{equation}
		\begin{aligned}
			\frac{dx_g}{dt} =\frac{\delta}{2} f_g x_g(1-x_g)[w(x_g)-C_g] .
		\end{aligned}
	\end{equation}
	In this case, the replicator dynamics are recovered.
	Setting $\frac{dx_g}{dt}=0$, for a relative vaccination cost $0 < C_g < w(0)$, we can compute that the system converges to the stable interior equilibrium $\hat{x_g} = 1 + \frac{\mathrm{ln}(1 - C_g)}{C_g\mathscr{R}_0}$, which is equal to the Nash equilibrium $x^*$ in Eq.~(\ref{Nash}). Therefore, weak intensity of selection (low  $\delta$ values) leads to the equilibrium vaccination level under imitation dynamics converging to the Nash equilibrium. 

	$\bullet$ Case 2: The intensity of selection $\delta \rightarrow \infty$. Consider the special case of $C_g \rightarrow 0$, so we have $-\frac{\delta}{2}C_g \rightarrow 0 $.  There are $\mathrm{tanh}(-\frac{\delta}{2}C_g) \sim -\frac{\delta}{2}C_g$ and $\mathrm{tanh}[\frac{\delta}{2}(1-C_g)] \sim 1$.
	The Eq.~(\ref{eq.20}) simplifies to
	\begin{equation}
		\begin{aligned}
			\frac{dx_g}{dt} =f_g x_g(1-x_g)\big(-\frac{\delta}{2}C_g[1-w(x_g)]+w(x_g)\big) .
		\end{aligned}
	\end{equation}
	Setting $\frac{dx_g}{dt}=0$, the stable interior equilibrium is calculated as $\hat{x}=1-\left(1+\frac{\delta C_g }{2}\right) \ln \left(1+\frac{\delta C_g}{2}\right) /\left(\frac{\delta C_g}{2} \mathscr{R}_0\right)$. When $C_g \rightarrow 0$, $\hat{x} \approx 1-\frac{1+\frac{\delta C_g}{4}}{\mathscr{R}_0}$ and Nash equilibrium $x_g^* \approx 1-\frac{1+\frac{ C_g}{2}}{\mathscr{R}_0}$. Therefore, when $\delta \rightarrow \infty$ and $C_g \rightarrow 0$, we have $\hat{x_g}<x_g^*$, indicating that when relative vaccination costs are low, strong intensity of selection (large $\delta$ values) leads to vaccination levels dropping below the Nash equilibrium.

\section{ Numerical simulations}\label{numerical} 
In this section, based on a data-driven model (see the electronic supplementary materials-C) incorporating the age structure and contact matrices, we implement the age-based epidemic-game coevolution model on the simplicial complex. 

\begin{figure*}[h]
	\centering
	\includegraphics[scale=1.18]{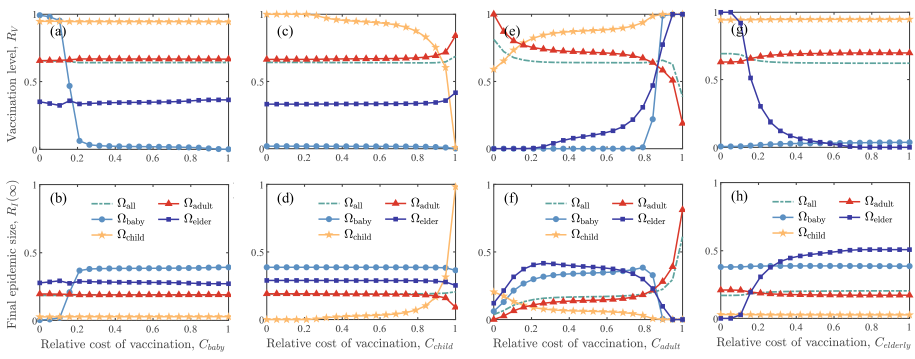}
	\caption{The effect of the relative vaccination cost $C_g$ for $\Omega_{g}$ on the vaccination level ($R_V$) and final epidemic size ($R_I(\infty)$) of five age level $\Omega_l$, where $l \in \rm{\{baby, child, adult, elderly, all\}}$ with $g =\rm{baby}$ (a)(b), $g =\rm{child}$ (c)(d), $g =\rm{adult}$ (e)(f), and $g =\rm{elderly}$ (g)(h). Different colors represent the five age levels corresponding to experimental results $R_V$ and $R_I(\infty)$. In (a)(b), the vaccination level $R_V$ of group $\Omega_{\rm{baby}}$ rapidly decreases with a slight increase in relative vaccination cost $C_{\rm{baby}}$; in (c)(d), the vaccination level $R_V$ of group $\Omega_{\rm{child}}$ remains relatively high, regardless of changes in relative vaccination cost $C_{\rm{child}}$; in (e)(f), the vaccination level $R_V$ of group $\Omega_{\rm{adult}}$ exhibits a rotating ``S" shape closely tied to variations in relative vaccination cost $C_{\rm{adult}}$; in (g)(h), the vaccination level $R_V$ of group $\Omega_{\rm{elderly}}$ gradually decreases with an increase in relative vaccination cost $C_{\rm{elderly}}$. These experimental phase transition results reflect the vaccination characteristics of different age groups, summarized in Table~\ref{table2}. For each experiment, the horizontal coordinate $C_g\in [0,1]$ and the relative vaccination costs for the remaining groups are the default costs presented in  supplementary materials-C. For example, in (a)(b), $C_{\rm{baby}}\in [0,1]$ and the costs of the remaining groups are $C_{\rm{child}}=C^{def}_{\rm{child}}$, $C_{\rm{adult}}=C^{def}_{\rm{adult}}$ and $C_{\rm{elderly}}=C^{def}_{\rm{elderly}}$, respectively. Other parameters: $I_0=1e-5$, $\delta=10$, $\beta=0.3$, $\beta^{\Delta}=0.4$, and $\gamma=0.8$.}
	\label{fig2}
\end{figure*}

The numerical experiments are conducted to investigate how relative vaccination costs, age structure, and the reinforcement infection effect influence vaccination behavior and infectious disease transmission. We utilize $R_V$ to represent the extent of vaccine uptake in the population, indicating the level of vaccination. $R_I(\infty)$ represents the final epidemic size. Both $R_V$ and $R_I(\infty)$ are discussed across five age levels in the population, denoted as $\Omega_l$, where $l \in \rm{\{baby, child, adult, elderly, all\}}$. The $\Omega_{\rm{all}}$ represents the entire population across age structures, while the other four age levels correspond to the age structures. The experimental results in the figures characterize the level of vaccine uptake and the extent of infectious disease transmission within each group in $\Omega_l$, where $l \in \rm{\{baby, child, adult, elderly, all\}}$.
The initial infected population is randomly distributed across the entire population \cite{Sardar2022allee}, and $I_0=1e-5$ is set to represent the proportion of initially infected individuals. After undergoing 2800 seasonal influenza outbreaks, corresponding to 2800 iterations of the vaccination game, the changes in individual vaccination strategies gradually stabilize and the evolutionary system is at equilibrium. All figs.~\ref{fig2}--\ref{fig7} depict the experimental results at equilibrium. In general, the numerical experiments are divided into the following discussion aspects.
\subsubsection{Population vaccination characteristics}\label{Popu vac chara}
\begin{table*}[htp]
	\centering
	\caption{Population vaccination characteristics.}
	\label{table2}
	\begin{tabular}{ccc}
		\toprule
		Age group & Vaccination characteristics & Reason description  \\
		\midrule
		\multirow{2}{*}{$\Omega_{\rm{baby}}$} &  \multirow{2}{*}{absolutely conservative} &  Once the relative cost of the vaccination is slightly higher, \\
		&  &  individuals within the $\Omega_{\rm{baby}}$ give up on vaccination.\\
		\multirow{2}{*}{$\Omega_{\rm{child}}$} &  \multirow{2}{*}{vaccination-active} &  As long as there is a risk of infection, individuals within \\
		&  &   the  $\Omega_{\rm{child}}$ choose to get vaccinated regardless of the vaccination cost.\\
		\multirow{2}{*}{$\Omega_{\rm{adult}}$} & \multirow{2}{*}{cost-oriented}  &  The vaccination level in the $\Omega_{\rm{adult}}$ is strongly correlated with \\
		&  &   changes in relative vaccination cost, showing an ``S" shaped pattern.\\
		\multirow{2}{*}{$\Omega_{\rm{elderly}}$} & \multirow{2}{*}{relatively conservative} &  The slight increase in the relative cost of vaccination leads to  \\
		&  &   individuals within $\Omega_{\rm{elderly}}$ gradually giving up on vaccination.\\
		\bottomrule
	\end{tabular}
\end{table*}

Taking into account the age-stratified population data of Beijing, China, we explore the vaccination characteristics of the population in this city. Based on the population age structure, we plotted the variation of vaccination level ($R_V$) and final epidemic size ($R_I(\infty)$) in relation to the relative vaccination costs for different age groups in Fig.~\ref{fig2}. Based on the analysis of various phase transition phenomena, the vaccination characteristics of the four age groups are defined as shown in Table~\ref{table2}.

Group $\Omega_{\rm{baby}}$ is characterized as an absolutely conservative type of vaccination behavior. 
As shown in Fig.~\ref{fig2}(a) and (b), we observe that as the relative vaccination cost $C_{\rm{baby}}$ increases, the vaccination level within group $\Omega_{\rm{baby}}$ sharply decreases and tends towards zero, leading to an increase in the final epidemic size within the $\Omega_{\rm{baby}}$ group. This phenomenon indicates that individuals within the $\Omega_{\rm{baby}}$ group tend to have a consistent acceptance level towards the relative vaccine cost and there exists an acceptable cost threshold. Once the vaccination cost surpasses this threshold, where the infection cost is lower than the threshold, individuals within the $\Omega_{\rm{baby}}$ group abandon vaccination. 
Therefore, the Vaccination characteristic of group $\Omega_{\rm{baby}}$ is defined as an \emph{absolutely conservative type} and exhibiting a prominent free-riding mentality. Specifically, the vaccine characteristics of group $\Omega_{\rm{baby}}$ actually describe the characteristics of parents towards vaccinating the group $\Omega_{\rm{baby}}$.

Group $\Omega_{\rm{child}}$ is characterized as a vaccination-active type of vaccination behavior. The ``vaccination-active" implies that group  $\Omega_{\rm{child}}$ is highly proactive about vaccination, i.e., individuals in the $\Omega_{\rm{child}}$ group choose to get vaccinated whenever there is a risk of infection, regardless of the cost. As shown in Fig.~\ref{fig2}(c) and (d), we find that the vaccination level in the $\Omega_{\rm{child}}$ group remained consistently high until the relative vaccination cost $C_{\rm{child}}$ reached a higher value (around 0.8 in the figure). However, once the relative vaccination cost $C_{\rm{child}}$ exceeded 0.8, the vaccination level in the $\Omega_{\rm{child}}$ group rapidly declined to zero. This phenomenon indicates that the $\Omega_{\rm{child}}$ group has a high acceptance level towards the vaccination cost and can be characterized as being \emph{vaccination-active}. Specifically, as long as there is an infection risk, individuals in the $\Omega_{\rm{child}}$ group choose to get vaccinated. 
Considering the impact of $C_{\rm{child}}$ on other age groups, we find that as the $C_{\rm{child}}$ in the $\Omega_{\rm{child}}$ group increases to a sufficiently high level, the vaccination level in other age groups begins to show an upward trend. This is due to the extensive social interactions between the $\Omega_{\rm{child}}$ group and other age groups. 

Group $\Omega_{\rm{adult}}$ is characterized as a cost-oriented type of vaccination behavior. As shown in Fig.~\ref{fig2}(e) and (f), we find that as the relative vaccination cost $C_{\rm{adult}}$ increases in the $\Omega_{\rm{adult}}$ group, the vaccination level in the $\Omega_{\rm{adult}}$ group and the overall population tend to align, following a rotated ``S" shape. In other words, as the relative vaccination cost $C_{\rm{adult}}$ increases, the vaccination level initially decreases gradually, then stabilizes, and finally decreases to a lower value. This indicates that the vaccination level in the $\Omega_{\rm{adult}}$ group is closely related to changes in the $C_{\rm{adult}}$ and can be characterized as being \emph{cost-oriented}. Specifically, we discuss the vaccination and infection phenomena in the $\Omega_{\rm{adult}}$ across three stages. 
(i) In the initial stage, $[0, 0.2)$: As the $C_{\rm{adult}}$ increases, the vaccination level in the $\Omega_{\rm{adult}}$ group and the overall population gradually decrease, while the final epidemic size increases. The $R_V$ in the $\Omega_{\rm{child}}$ group increases, leading to a decrease in the final epidemic size. Also, the vaccination level in the $\Omega_{\rm{baby}}$ and $\Omega_{\rm{elderly}}$ groups remain at zero, and the final epidemic size significantly expands.
(ii) In the stable stage, $[0.2, 0.8)$: The vaccination level in all four groups and the overall population remain relatively stable. The vaccination level in the $\Omega_{\rm{adult}}$ and $\Omega_{\rm{child}}$ groups stabilize at a higher value, while the vaccination level in the $\Omega_{\rm{baby}}$ group remains at zero. Only the vaccination level in the $\Omega_{\rm{elderly}}$ group increases.
(iii) In the final stage, $[0.8, 1]$: The vaccination level in the $\Omega_{\rm{adult}}$ group and the overall population decrease to a smaller value, and the final epidemic size reaches a larger value. Meanwhile, the vaccination level in the $\Omega_{\rm{child}}$, $\Omega_{\rm{baby}}$, and $\Omega_{\rm{elderly}}$ groups sharply increase.

Group $\Omega_{\rm{elderly}}$ is characterized as a relatively conservative type of vaccination behavior. Observing Fig.~\ref{fig2}(g) and (h), we find that as the relative vaccination costs $C_{\rm{elderly}}$ increase, similar to the changes in $\Omega_{\rm{baby}}$ shown in Fig.~\ref{fig2}(a) and (b), the vaccination level in $\Omega_{\rm{elderly}}$ starts to decline at a relatively low value of $C_{\rm{elderly}}$. This indicates that both groups exhibit conservative vaccination behavior and attempt to free-ride. However, unlike the sharp decrease in vaccination levels observed in $\Omega_{\rm{baby}}$, the vaccination level in $\Omega_{\rm{elderly}}$ gradually declines, leading to an increase in the final epidemic size within the $\Omega_{\rm{elderly}}$ group. This suggests that the $\Omega_{\rm{elderly}}$ group exhibits heterogeneity in accepting vaccination costs. Individuals in $\Omega_{\rm{elderly}}$ have varying levels of acceptance towards vaccination costs, and they only refuse vaccination when the costs exceed their personal threshold. Therefore, the vaccination characteristics of $\Omega_{\rm{elderly}}$ is defined as \emph{relative conservative type}. Additionally, the increase in $C_{\rm{elderly}}$ has a slight positive impact on the vaccination level in the $\Omega_{\rm{adult}}$ group, while it does not affect the other groups. 
\subsubsection{Impact of higher-order interaction in household}
\begin{figure}[h]
	\centering
	\includegraphics[scale=0.92]{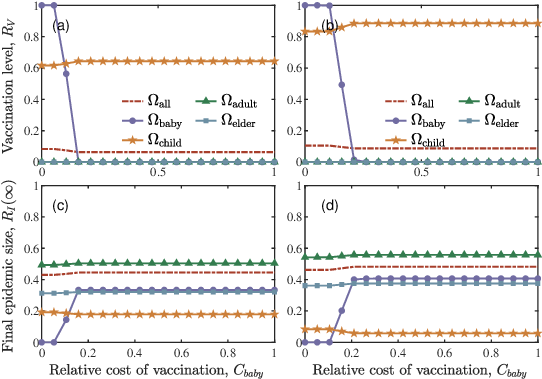}
	\caption{The Vaccination level $R_V$ and final epidemic size $R_I(\infty)$ of five age level $\Omega_l$, $l \in \rm{\{baby, child, adult, elderly, all\}}$ versus the relative cost of vaccination $C_{\rm{baby}}$ for $\Omega_{\rm{baby}}$ with $\beta^{\Delta}=0.1$ (a)(c) and  $\beta^{\Delta}=0.95$ (b)(d). Different colors correspond to the experimental results for different age groups. The comparative analysis of experimental results in (a)(c) and (b)(d) indicates that the increase in reinforcement infection effect $\beta^{\Delta}$ leads to an expansion of the relative vaccination cost at which the group $\Omega_{\rm{baby}}$ refuses vaccination, while the vaccination level $R_V$ of group $\Omega_{\rm{child}}$ expands. Other parameters: $I_0=1e-5$, $C_{\rm{child}}=C^{def}_{\rm{child}}$, $C_{\rm{adult}}=C^{def}_{\rm{adult}}$, $C_{\rm{elderly}}=C^{def}_{\rm{elderly}}$, $\delta=10$, $\beta=0.1$, and $\gamma=0.8$.}
	\label{fig3}
\end{figure}
The high-order interactions in the household scenario bring about a reinforcement infection effect $\beta^{\Delta}$, which depicts the collective infection phenomenon occurring among family members in reality. We demonstrate the influence of high-order interactions on vaccination behavior and infectious diseases transmission by adjusting the parameters of $\beta^{\Delta}$. For $\Omega_{\rm{baby}}$, in Fig.~\ref{fig3}, we find that an increase in the $\beta^{\Delta}$, leads to a higher relative vaccination cost for the population $\Omega_{\rm{baby}}$ to refuse vaccination. This is because the $\beta^{\Delta}$ increases the infection risk for the population $\Omega_{\rm{baby}}$, causing them to refuse vaccination only at higher costs. Additionally, as $C_{\rm{baby}}$ increases, when the vaccination level for the population $\Omega_{\rm{baby}}$ decreases to zero, the higher-order infection rate increases, resulting in an expansion of the final epidemic size for the group $\Omega_{\rm{baby}}$. Furthermore, the increase in $\beta^{\Delta}$ leads to an expansion of the vaccination level and a reduction in the infection size for the group $\Omega_{\rm{child}}$. This indicates that the group $\Omega_{\rm{child}}$ has extensive contact with other household members, which increases their infection risk, thereby increasing the vaccination level and reducing the number of infections. 

\begin{figure}[h]
	\centering
	\includegraphics[scale=0.92]{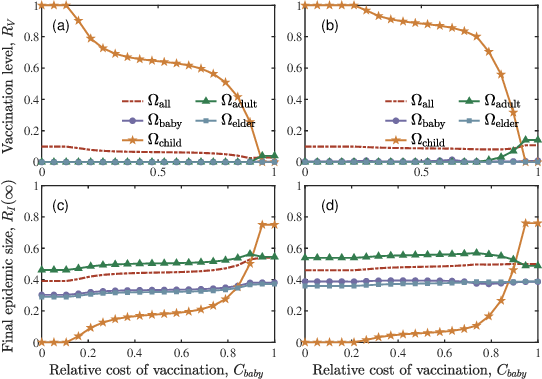}
	\caption{The influence of the relative cost of vaccination $C_{\rm{child}}$ for $\Omega_{\rm{child}}$ on the vaccination level ($R_V$) and final epidemic size ($R_I(\infty)$) of five age level $\Omega_l$, where $l \in \rm{\{baby, child, adult, elderly, all\}}$ with $\beta^{\Delta}=0.1$ (a)(c) and  $\beta^{\Delta}=0.95$ (b)(d). The experimental results in (a)(c) and (b)(d) demonstrate that the rising reinforcement infection effect $\beta^{\Delta}$ promotes an increase in the vaccination level $R_V$ of groups $\Omega_{\rm{child}}$ and $\Omega_{\rm{adult}}$.
		Other parameters are consistent with Fig.~\ref{fig3}.}
	\label{fig4}
\end{figure}
\begin{figure}[b]
	\centering
	\includegraphics[scale=0.92]{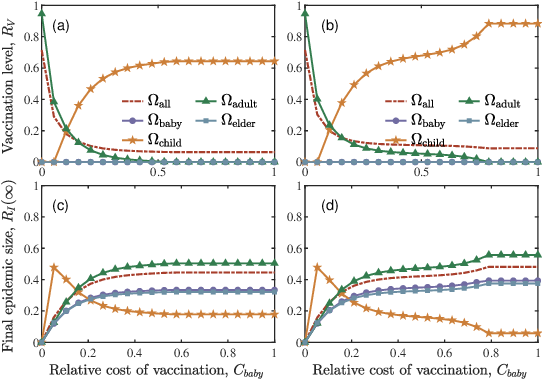}
	\caption{Effects of the relative cost of vaccination $C_{\rm{adult}}$ on vaccination game with $\beta^{\Delta}=0.1$ (a)(c) and  $\beta^{\Delta}=0.95$ (b)(d). The comparative analysis of experimental results in (a)(c) and (b)(d) indicates that the increase in the reinforcement infection effect $\beta^{\Delta}$ results in the vaccination level $R_V$ of the group $\Omega_{\rm{adult}}$ decreasing to zero only when the relative vaccination cost expands to around $0.8$. This demonstrates the enhancement of the vaccination willingness in the group $\Omega_{\rm{adult}}$.
		Other parameters are consistent with Fig.~\ref{fig3}.}
	\label{fig5}
\end{figure}
Next, for $\Omega_{\rm{child}}$, as shown in Fig.~\ref{fig4}, we find that an increase in $\beta^{\Delta}$ leads to an expansion of the vaccination level for $\Omega_{\rm{child}}$. Additionally, when $C_{\rm{child}}$ is high, the elevated $\beta^{\Delta}$ in the family promotes an increase in the vaccination level for $\Omega_{\rm{child}}$ and $\Omega_{\rm{adult}}$. This is because when $C_{\rm{child}}$ is high, the vaccination level for $\Omega_{\rm{child}}$ decreases significantly, resulting in a higher number of infections within this group. Given the extensive contact between $\Omega_{\rm{child}}$ and $\Omega_{\rm{adult}}$ within the family and the elevated reinforcement infection effect, the infection risk for $\Omega_{\rm{adult}}$ in the family increases, thereby promoting vaccination. Furthermore, due to the lower contact intensity between $\Omega_{\rm{baby}}$ and $\Omega_{\rm{child}}$ as well as between $\Omega_{\rm{baby}}$ and $\Omega_{\rm{elderly}}$ within the family, the infection risk is low, resulting in a vaccination level of zero for both $\Omega_{\rm{baby}}$ and $\Omega_{\rm{elderly}}$.

For $\Omega_{\rm{adult}}$, as shown in Fig~\ref{fig5}, we observe that an increase in the reinforcement infection effect $\beta^{\Delta}$ expands the cost range for which the vaccination disappears for the group $\Omega_{\rm{adult}}$ to around 0.8. This indicates that the reinforcement infection effect within households increases the infection risk for the group $\Omega_{\rm{adult}}$, leading to an expansion of the cost range for vaccination disappearance. Besides, the increased $\beta^{\Delta}$ promotes a significant expansion of the vaccination level for the group $\Omega_{\rm{child}}$. This is because the group $\Omega_{\rm{child}}$ has extensive contact with other household members, such as parents, and their infection probability is largely influenced by them. 
After $C_{\rm{adult}}>0.8$, an increase in $\beta^{\Delta}$ leads to an enlargement of the final epidemic size for the group $\Omega_{\rm{baby}}$, $\Omega_{\rm{adult}}$, and $\Omega_{\rm{elderly}}$. However, due to the substantial increase in the vaccination rate of the group $\Omega_{\rm{child}}$, which is sufficient to control the spread of the infectious diseases, the final epidemic size for the group $\Omega_{\rm{child}}$ decreases significantly.

\begin{figure}[h]
	\centering
	\includegraphics[scale=0.92]{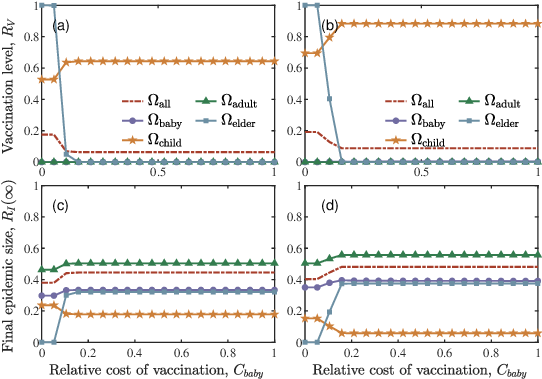}
	\caption{Phase diagram for the influence of relative cost of vaccination $C_{\rm{elderly}}$ on the vaccination game with $\beta^{\Delta}=0.1$ (a)(c) and  $\beta^{\Delta}=0.95$ (b)(d). Comparing the experimental results in (a)(c) and (b)(d), it is observed that the enhanced reinforcement infection effect $\beta^{\Delta}$, to some extent, led to the group $\Omega_{\rm{adult}}$ refusing vaccination at a higher relative vaccination cost $C_{\rm{elderly}}$, slightly enhancing the vaccination willingness of $\Omega_{\rm{adult}}$. 
		Other parameters are consistent with Fig.~\ref{fig3}.}
	\label{fig6}
\end{figure}
In the final experiment of this section, as shown in Fig.~\ref{fig6}, we find that an increase in the $\beta^{\Delta}$ raises the cost threshold for the group $\Omega_{\rm{elderly}}$ to refuse vaccination. This indicates that the increase in the $\beta^{\Delta}$ enhances the infection risk for the group $\Omega_{\rm{elderly}}$ within households. Besides, an increase in $\beta^{\Delta}$ expands the vaccination level for the group $\Omega_{\rm{child}}$ and reduces the final epidemic size. 

\subsubsection{Generalizability of vaccination characteristics}
\begin{figure*}[h]
	\centering
	\includegraphics[scale=1.18]{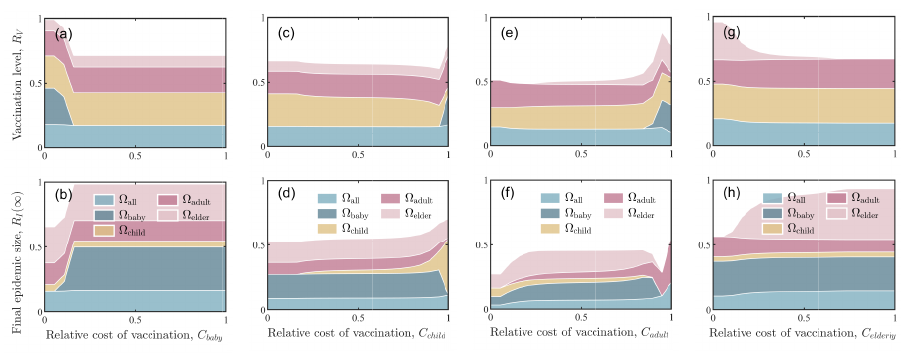}
	\caption{The impact of the relative vaccination cost on the vaccination game based on the age-stratified population data from New York. The different colored areas represent the vaccination levels $R_V$ (a)(c)(e)(g) and final epidemic sizes $R_I(\infty)$ (b)(d)(f)(h) corresponding to the population levels $\Omega_{l}$, where $l \in \rm{\{baby, child, adult, elderly, all\}}$. Specifically, the light blue represents $\Omega_{\rm{all}}$, dark blue represents $\Omega_{\rm{baby}}$, yellow represents $\Omega_{\rm{child}}$, dark pink represents $\Omega_{\rm{adult}}$, and light pink represents $\Omega_{\rm{elderly}}$. Comparing the experiments conducted based on the population data of Beijing, China, as shown in Fig.~\ref{fig2}, consistent experimental results are observed. This confirms the generalizability of age-based vaccination characteristics across different populations. The experimental parameters are consistent with Fig.~\ref{fig2}.}
	\label{fig7}
\end{figure*}
The vaccination characteristics of the four age groups are summarized in Sec~\ref{Popu vac chara}. To verify the universality of age-based vaccination characteristics across different populations, we attempt to replicate the experiments in Sec~\ref{Popu vac chara} using age-stratified data from another country \cite{Mistry2021Inferring}. Therefore, relying on population contact and age-stratified data from the United States, New York, we once again plot the changes in vaccination level $R_V$ and final epidemic size $R_I(\infty)$ with respect to the relative vaccination cost $C_g$ for $\Omega_{g}$, where $g \in\rm{\{baby, child, adult, elderly\}}$. 

The Fig.~\ref{fig7} presents the experimental results in a stacked graph, utilizing the perspective of ``proportions of areas". 
Firstly, the population group $\Omega_{\rm{child}}$ exhibits the highest vaccination level. We observe that the yellow region, representing the area of $\Omega_{\rm{child}}$ vaccination level, has the largest proportion in subfigures (a)(c)(e)(g), indicating the highest vaccination level among $\Omega_{\rm{child}}$. 
Secondly, the dark blue region has the lowest proportion in (a)(c)(e)(g), indicating that $\Omega_{\rm{baby}}$ has the lowest vaccination level. This may be attributed to the relatively limited social contacts of $\Omega_{\rm{baby}}$ compared to other groups, such as going to school, work, or parks, resulting in higher infection risks and higher vaccination levels for the other three groups. 
Thirdly, both $\Omega_{\rm{baby}}$ and $\Omega_{\rm{elderly}}$ show the largest final epidemic size, as evidenced by the larger proportion of light pink and dark blue regions in (b)(d)(f)(h). This is because the vaccination characteristics of both $\Omega_{\rm{baby}}$ and $\Omega_{\rm{elderly}}$ is conservative, and their low vaccination levels lead to a wider range of infections. Finally, when the number of individuals vaccinated in $\Omega_{\rm{child}}$ or $\Omega_{\rm{adult}}$ starts to increase, we observe a decrease in the vaccination levels of other groups in (c)(e). This phenomenon reflects the free-rider mentality of $\Omega_{\rm{baby}}$ and $\Omega_{\rm{elderly}}$, where they attempt to avoid personal vaccination but rely on others' vaccination to achieve herd immunity and protect themselves from infection. In general, these observations demonstrate that the experimental results are consistent with the results shown in Fig.\ref{fig2}, confirming the generalizability of age-based vaccination characteristics to different populations.

\section{Conclusions} \label{conclusion}
This work constructs an epidemic-game coevolution model that considers age structure and higher-order interactions simultaneously.
The mean-field approach is extended to analyze the dynamic equilibrium and ESSs based on imitation dynamics in the well-mixed population. 
Extensive numerical experiments explore the impact of age-structured vaccination characteristics and evolution, as well as the influence of higher-order interactions on vaccination behavior. The vaccination of $\Omega_{\rm{baby}}$ is described as a group of absolutely conservative type, the vaccination profile of $\Omega_{\rm{child}}$ vaccination-active type, $\Omega_{\rm{child}}$ is classified as a cost-oriented type, and group $\Omega_{\rm{elderly}}$ is categorized as a relatively conservative type. Based on population data from New York, a vaccination game demonstrated the generalizability of age-structured vaccination characteristics.
Additionally, the vaccination level is lowest in the $\Omega_{\rm{baby}}$ group, while the $\Omega_{\rm{child}}$ group has the highest vaccination level. 
The reinforcement infection effect resulting from higher-order interactions leads to an increase in the vaccination levels of the $\Omega_{\rm{child}}$ and $\Omega_{\rm{adult}}$. 



Overall, our research findings demonstrate significant differences in behaviors towards vaccination across age groups, highlighting the need for more precise and personalized vaccine promotion strategies. Additionally, our study offers a theoretical foundation and a new perspective for future prediction and intervention of vaccination behavior. 
However, this study lacks consideration of a multi-city context, which may limit the model's applicability to more complex, real-world scenarios. The mobility and the imitation dynamics between different cities could significantly impact vaccination decisions and epidemic dynamics \cite{Chang2020Impact,Chang2019Effects}. Future research could incorporate multi-city SIR modeling to further extend the current findings.

\section*{Code availability}
The code can be publicly accessed at the dedicated online repository \url{https://github.com/nienianni/epidemic-game-coevolution.git}


\bibliographystyle{IEEEtr}
\bibliography{mybibfile}

\vfill

\end{document}